\documentclass[aps,prl,twocolumn,amsmath,amssymb,nofootinbib,showpacs,superscriptaddress]{revtex4-1}
\usepackage[english]{babel}
\usepackage{latexsym}
\usepackage{graphics}
\usepackage{graphicx}
\usepackage{epsfig}
\usepackage{color}
\usepackage{bm}
\usepackage{amsmath}
\usepackage{amssymb}
\usepackage{amsthm}
\usepackage{dcolumn}
\usepackage{bm}
\usepackage{float}
\usepackage{hyperref}
\usepackage{color}
\usepackage{epstopdf}
\usepackage{cleveref}
\usepackage[svgnames]{xcolor}

\usepackage{soul} 

\newcommand{\Mod}[1]{\ (\mathrm{mod}\ #1)}
\hypersetup{hidelinks,colorlinks=true,allcolors=DarkBlue}

\theoremstyle{remark}

\begin{document}

\preprint{APS/123-QED}

\title{Hybrid quantum-classical machine learning for generative chemistry and drug design}

\author{A.I. Gircha}
\affiliation{Russian Quantum Center, Skolkovo, Moscow 143025, Russia}

\author{A.S. Boev}
\affiliation{Russian Quantum Center, Skolkovo, Moscow 143025, Russia}

\author{K. Avchaciov}
\affiliation{Gero PTE, 60 Paya Lebar Road \# 05-40B, Paya Lebar Square, 409051, Singapore}

\author{P.~O.~Fedichev}
\email[Correspondence and requests for materials should be addressed to A. K. Fedorov and P.O. Fedichev ]{(akf@rqc.ru, peter.fedichev@gero.ai)}
\affiliation{Gero PTE, 60 Paya Lebar Road \# 05-40B, Paya Lebar Square, 409051, Singapore}

\author{A.K. Fedorov}
\email[Correspondence and requests for materials should be addressed to A. K. Fedorov and P.O. Fedichev ]{(akf@rqc.ru, peter.fedichev@gero.ai)}
\affiliation{Russian Quantum Center, Skolkovo, Moscow 143025, Russia}

\date{\today}
\begin{abstract}
Deep generative chemistry models emerge as powerful tools to expedite drug discovery. 
However, the immense size and complexity of the structural space of all possible drug-like molecules pose significant obstacles, which could be overcome with hybrid architectures combining quantum computers with deep classical networks. 
As the first step toward this goal, we built a compact discrete variational autoencoder (DVAE) with a Restricted Boltzmann Machine (RBM) of reduced size in its latent layer. 
The size of the proposed model was small enough to fit on a state-of-the-art D-Wave quantum annealer and allowed training on a subset of the ChEMBL dataset of biologically active compounds. 
Finally, we generated $2331$ novel chemical structures with medicinal chemistry and synthetic accessibility properties in the ranges typical for molecules from ChEMBL. 
The presented results demonstrate the feasibility of using already existing or soon-to-be-available quantum computing devices as testbeds for future drug discovery applications.
\end{abstract}

\maketitle
\section{Introduction}

Drug design is the process of identifying biologically active compounds and relies on the efficient generation of novel, drug-like, and yet synthetically accessible compounds. 
So far, only about $10^8$ substances have ever been synthesized~\cite{kim2016pubchem}, whereas the total number of realistic drug-like molecules is estimated to be in the range between $10^{23}$ and $10^{60}$~\cite{polishchuk2013estimation}. 
This is why deep learning~\cite{deng2022artificial} and particularly deep generative models~\cite{gomez2018automatic, guimaraes2017objective, griffiths2020constrained, zhavoronkov2019deep} 
are believed to be helpful in generative chemistry and computation drug discovery applications involving sampling and scoring novel chemical structures from the very large and hitherto unknown distributions of possible drug-like molecules 
(see examples and benchmarks in Refs.~\cite{brown2019guacamol,polykovskiy2020molecular,paperswithcodeGENZINC}). 

A fully developed generative model should implicitly estimate the fundamental molecular properties, such as stability and synthetic accessibility for each generated compound and its intermediate products. 
All those features depend on the ability of the network architecture to approximate the solutions of the underlying quantum mechanical problems, which is computationally hard for molecules of realistic size. 
Quantum computers are naturally good for solving complex quantum many-body problems~\cite{Lloyd} and thus may be instrumental in applications involving quantum chemistry~\cite{McArdle,Bauer_2020,Aspuru-Guzik2018,Fedorov2021}. 
Moreover, quantum algorithms can speed up machine learning~\cite{Aspuru-Guzik2018,Biamonte2017}. 
Therefore, one can expect that quantum-enhanced generative models~\cite{Gao2021}, including quantum GANs~\cite{li2021quantum}, may eventually be developed into ultimate generative chemistry algorithms. 

\begin{figure}[h!]
	\includegraphics[width=0.75\linewidth]{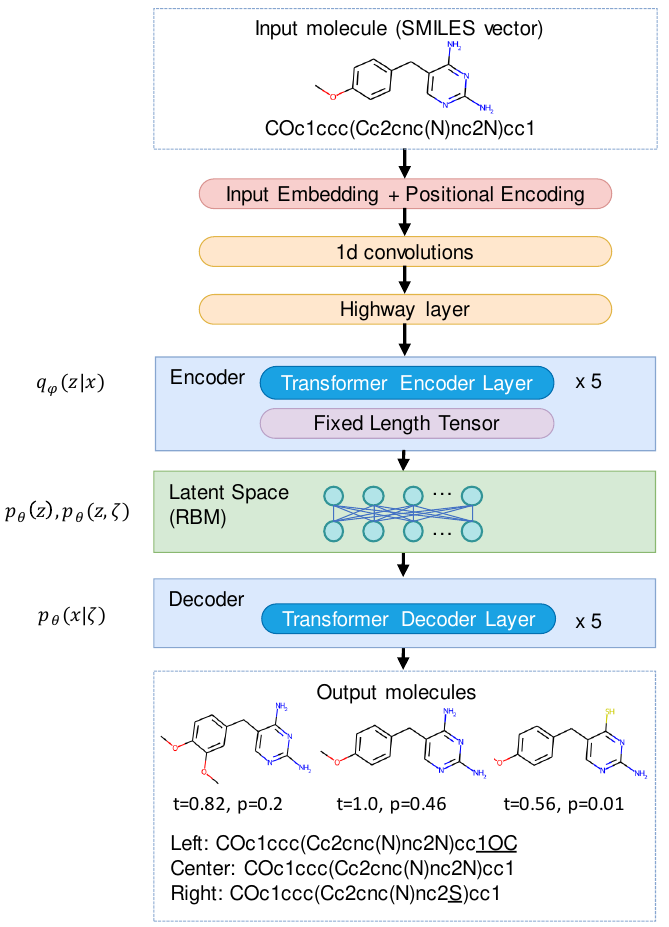}
	\vskip-3mm
	\caption{Scheme of the DVAE learning a joint probability distribution over the molecular structural features $x$ and their latent variable-representations (discrete $z$ and continuous $\zeta$). 
	Here, $q_\phi(z|x)$ and $p_\theta(x|\zeta)$ are the encoder and decoder distributions, respectively, whereas $p_\theta(z)$ is the prior distribution in the latent variable space and is approximated by RBM. 
	We provide an example of reconstruction of a target molecule (diaveridine) using Gibbs-300 model saved after 300 epochs of training (here $t \in [0,1]$ is the Tanimoto similarity between the initial molecule and its reconstruction, 
	$t=1.0$ corresponds to perfect reconstruction, $p$ is the output probability).}
	\label{fig:net-arch}
\end{figure}

Exploring the full potential of quantum machine-learning algorithms requires the development of fault-tolerant hardware~\cite{Biamonte2017}, which is not yet accessible. 
Meanwhile, readily available noisy intermediate-scale quantum (NISQ) devices~\cite{Preskill2018} provide a test-bed for the development and testing of quantum machine-learning algorithms for practical problems of modest size. 
For example, quantum annealing processors~\cite{Johnson2011} could potentially enable more efficient solving quadratic unconstrained binary optimization problems and approximating sampling from the thermal distributions of transverse Ising systems. 
These applications are attractive in the context of machine learning as tools both for solving optimization problems~\cite{Neven2008,Neven2012,Lidar2013,Lidar2017} and sampling~\cite{Andriyash2016,Perdomo-Ortiz2018,Melko2018,Dixit2021}.
Gate-based architectures are also of interest for machine learning~\cite{Biamonte2017}, 
in particular, in the context of quantum GANs, which are a subject of intensive research~\cite{Lloyd2018,Killoran2018,Aspuru-Guzik2019,Woerner2019,Pan2021} 
including recent demonstration of learning and generation of hand-written digit images on a quantum processor~\cite{Pan2021}. 

In this work, we prototyped a discrete variational autoencoder (DVAE, see Ref.~\cite{khoshaman2018quantum}), 
whose latent generative process is implemented in the form of a Restricted Boltzmann Machine (RBM) of a small enough size to fit readily available annealers. 
We trained the network on D-Wave annealer and generated $2331$ novel chemical structures with medicinal chemistry and synthetic accessibility properties in the ranges typical for molecules from ChEMBL. 
Hence, we demonstrated that the hybrid architecture might allow practical machine learning applications for generative chemistry and drug design. 
Once the hardware matures, the RBM could be turned into Quantum Boltzman Machine (QBM), and the whole system might be transformed into a Quantum VAE (QVAE,~\cite{khoshaman2018quantum}) and sample from richer non-classical distributions.

\section{Results}

We proposed and characterized a generative model (see Fig.~\ref{fig:net-arch}) in the form of a combination of a Discrete Variational Autoencoder (DVAE) model 
with a Restricted Boltzmann Machine (RBM) in the latent space~\cite{khoshaman2018quantum, rolfe2016discrete} and the Transformer model~\cite{vaswani2017attention}. 
The model learns good representations of chemical structures from ChEMBL, which is the manually curated database of biologically active molecules with drug-like properties~\cite{gaulton2017chembl}. 

Following Ref.~\cite{gomez2018automatic}, we used common SMILES~\cite{weininger1988smiles} 
encoding for organic molecules and trained the system to encode and subsequently decode molecular representations via optimizing evidence lower bound (ELBO) for DVAE log-likelihood~\cite{khoshaman2018quantum}:
\begin{equation}
\begin{aligned}
	\mathcal{L} (\mathbf{x}, \boldsymbol{\theta}, \boldsymbol{\phi}) = \mathbb{E}_{q_{\boldsymbol{\phi}}(\boldsymbol{\zeta} | \mathbf{x})}[\log p_{\boldsymbol{\theta}}(\mathbf{x} | \boldsymbol{\zeta})] + \\
	- \beta D_{KL}(q_{\boldsymbol{\phi}}(\mathbf{z}|\mathbf{x}) || p_{\boldsymbol{\theta}}(\mathbf{z})).
\end{aligned}
\label{eq:elbo}
\end{equation}
Here, $\mathbb{E}$ denotes the expectation value, $D_{KL}$ is the Kullback–Leibler (KL) divergence, 
and $p_{\boldsymbol{\theta}}(\mathbf{z})$ is the prior distribution in the latent variable space and is encoded by RBM as in Ref.~\cite{khoshaman2018quantum} (see MM).
The two layers of RBM contain $128$ units each, an RBM of this size can be sampled on readily available quantum annealers. 
We used the spike-and-exponential transformation~\cite{khoshaman2018quantum} as a smoothing probability distribution between the discrete $\mathbf{z}$ 
and continuous $\boldsymbol{\zeta}$ variables and employed the standard reparameterization trick to avoid calculating derivatives over random variables. 

The respective encoder and decoder functions, $q_{\boldsymbol{\phi}}(\mathbf{z} | \mathbf{x})$ and $p_{\boldsymbol{\theta}}(\mathbf{x} | \boldsymbol{\zeta})$, 
are approximated by the deep neural networks with Transformer layers each depending on its own set of adjustable parameters $\boldsymbol{\phi}$ and $\boldsymbol{\theta}$. 
We modified the KL divergence term with $\beta=0.1$ to avoid posterior collapse~\cite{yan2020rebalancing}. 

\begin{figure}[]
	\includegraphics[width=0.95\linewidth]{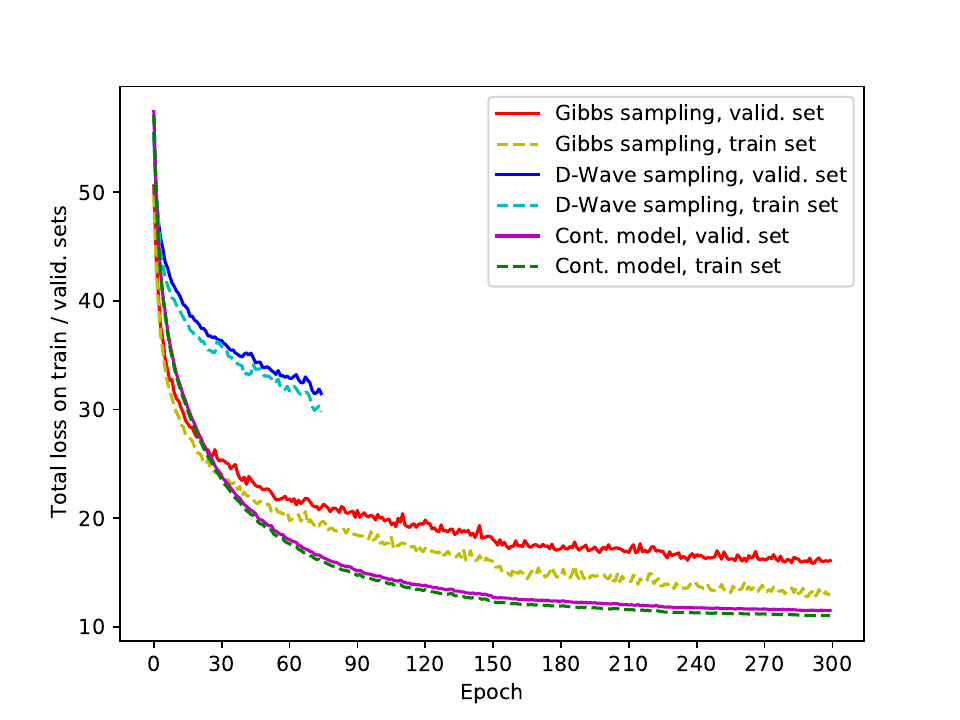}
	\caption{Learning curves of DVAE trained with classical Gibbs sampling (red, yellow) and samples from D-Wave annealer (blue, cyan). 
	Training on D-Wave suspended before reaching convergence due to resource limitation.
	Also, the learning curve of simpler model with continuous latent variables is shown (magenta, green).}
	\label{fig:loss}
\end{figure}

We trained the network for $300$ epochs until apparent convergence using Gibbs sampling (see the red and yellow lines in Fig.~\ref{fig:loss} representing the total loss over the validation and train sets, respectively). 
In what follows, we discuss the two checkpoints: the fully trained (Gibbs-300) and, for reference purposes, the intermediate model (Gibbs-75) appearing by the end of the $75$th epoch. 
We expect that with improvements in quantum hardware one can expect that (in particular, coherence times of qubits) training the DVAE with quantum annealing technique could be comparable or over overcome existing techniques.

VAE is a probabilistic model. In particular, this means that each of the discrete states in the latent variables is decoded into a probability distribution of SMILES-encoded molecules. 
On top of Fig.~\ref{fig:net-arch} we provide an example of encoding a particular molecule (diaveridine) and its reconstruction by the Gibbs-300 network (see the structures at the bottom). 
In this case, the target molecule was reconstructed exactly in $46\%$ runs (see the reconstruction probabilities and Tanimoto similarities to the target molecule next to the reported structures). 

\begin{figure} [h!]
	\includegraphics[width=1\linewidth]{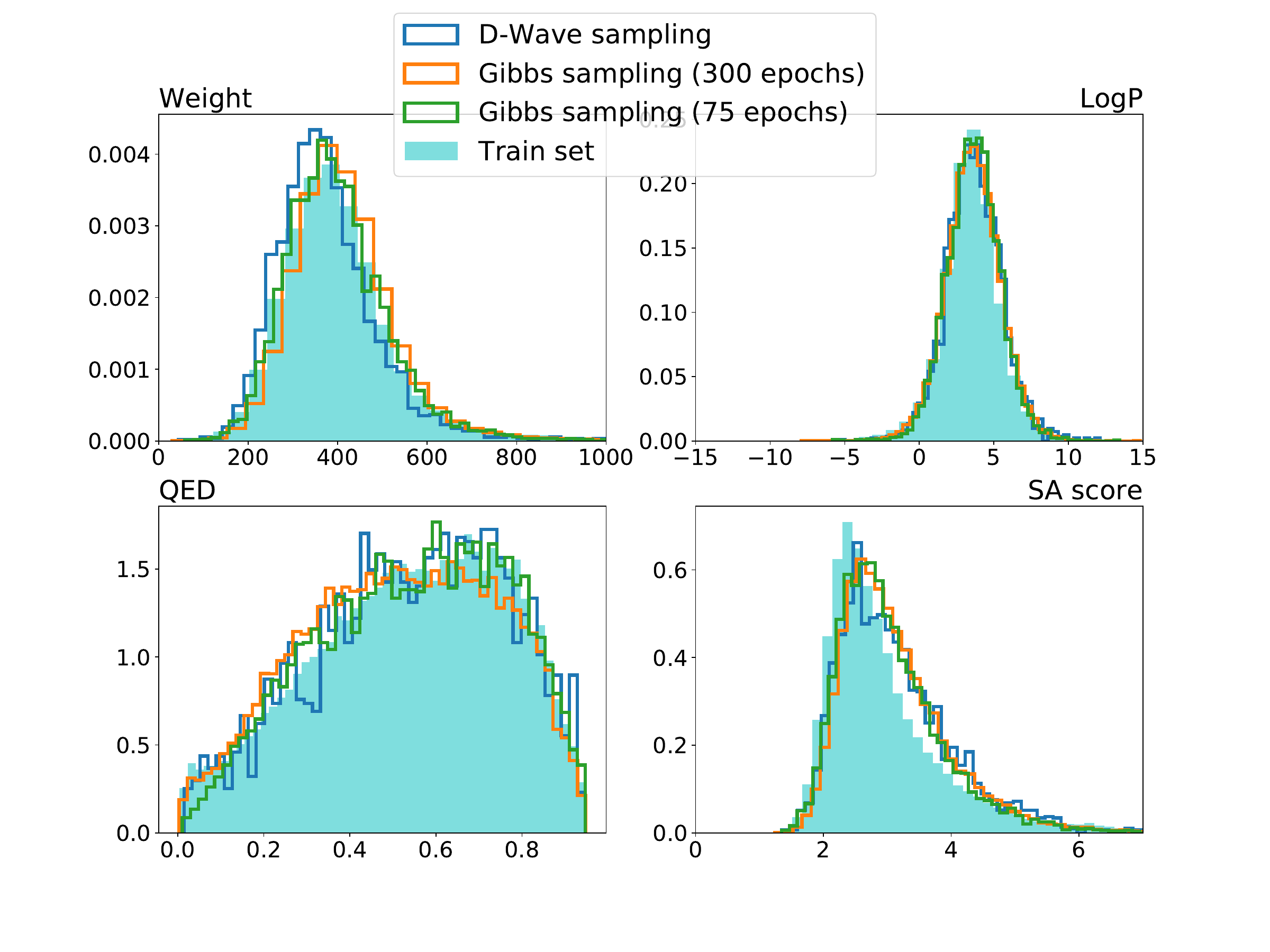}
	\vskip-8mm
	\caption{Distributions of  physico-chemical properties of the molecules produced by the proposed generative models discussed (same as in Table~\ref{table:metrics}). }
	\label{fig:comparison}
\end{figure}

DVAE is a generative model that can produce novel molecules with properties that presumably match those in the training set. 
In Fig.~\ref{fig:comparison} and Table~\ref{table:metrics} we compare the distributions of the basic biochemical properties of the molecules in the training set and among molecules generated by each of the models trained and discussed in this work. 
The novel molecules were mainly valid ($55\%$ and $69\%$ in Gibbs-75 ($10k$ molecules) and Gibbs-300 ($50k$ molecules) models, respectively). 
We kept track of molecular weight (MW), the water-octanol partition coefficient (logP), the synthetic accessibility (SA~\cite{Ertl2009}) score, and the quantitative estimation of drug-likeness (QED~\cite{Bickerton2012}) score, 
which are common physico-chemical properties for benchmarking molecular generative models~\cite{polykovskiy2020molecular}. 

Aside from the biochemical and drug-likeness properties, we also measured the novelty of generated molecules. 
Less than $1\%$ of the generated molecules ($0.36\%$ and $0.22\%$ in Gibbs-75 and Gibbs-300 models, respectively) had Tanimoto similarity larger than $0.9$ to any molecule in the training set and less than $10\%$
of the generated molecules are similar to any molecule in the training set with $T>0.7$ in both models. 
Extra training time improved both the validity of the generated molecules and brought the molecular properties closer to those found in the training set (see the relevant Gibbs-75 and Gibbs-300 columns in Table~\ref{table:metrics}).

The proposed network architecture is sufficiently compact to fit the D-Wave hardware. Hence, we were able to train the network using the annealer instead of Gibbs sampling. 
The learning of the hybrid model on D-Wave progressed slower than that on a classical computer using Gibbs sampling (see the blue solid and cyan dashed lines in Fig.~\ref{fig:loss} corresponding to the total loss of the model on the validation and the training sets). 
We had, however, to stop the training before reaching convergence at $75$th epoch due to the limited performance of the available quantum hardware. 
With its further improvements, we expect to have the ability to prolong the training.
Eventually, we used D-Wave to generate $4290$ molecular structures ($2331$ of which are grammatically correct, see Fig.~\ref{fig:comparison} and the corresponding column in Table~\ref{table:metrics}). 
As expected, the distributions of basic properties of the generated molecules were close to those obtained from the Gibbs-75 model and could be improved if more training time were available.

\begin{table}
\centering

\resizebox{\columnwidth}{!}{%
\begin{tabular}{ |c|c|c|c|c| } 
 \hline
     & Train set & \vtop{\hbox{\strut D-Wave sampl.}\hbox{\strut (75 epochs)}} & \vtop{\hbox{\strut Gibbs sampl.}\hbox{\strut (75 epochs)}} & \vtop{\hbox{\strut Gibbs sampl.}\hbox{\strut (300 epochs)}}\\ 
 \hline
 Number of samples & 153600 & 4290 & 10000 & 50000 \\
 MW    & 409.99 / 153.89 & 374.16 / 113.5 & 397.81 / 114.35 & 416.14 / 124.46\\ 
 LogP   & 3.41 / 2.01   & 3.68 / 1.9  & 3.64 / 1.77 & 3.61 / 1.87 \\
 QED   & 0.54 / 0.22   & 0.54 / 0.22  & 0.54 / 0.22 & 0.52 / 0.22 \\
 SAS   & 2.96 / 0.97   & 3.12 / 0.88  & 3.04 / 0.81 & 3.13 / 0.84\\
 Validity & 1.0  & 0.54  & 0.55  & 0.69 \\
 \hline
\end{tabular}%
}
\caption{
The parameters of distributions of physico-chemical properties of the molecules produced by the generative models discussed in this work. 
The entries in the table are mean/std values computed with the help of RDKit library~\cite{rdkitlib}, where MW is molecular weight, 
LogP is the octanol-water partition coefficient, 
QED is the quantitative estimation of drug-likeness, 
and SAS is the synthetic accessibility score. 
Validity is a fraction of good SMILES strings (which can be translated to a molecular graph) to a total number of generated strings.
}

\label{table:metrics}
\end{table}

\section{Discussion and outlook}

VAE are powerful generative machine learning models capable of learning and sampling from the unknown distribution of input data~\cite{kingma2013auto, kingma2019introduction}. 
As a first step towards building a hybrid quantum generative model, we prototyped the DVAE (along the lines of Ref.~\cite{khoshaman2018quantum}) with the RBM in its latent space~\cite{khoshaman2018quantum, rolfe2016discrete}.
If provided with a large dataset of drug-like molecules, such a system should learn implicit rules governing the stability and synthetic accessibility of small molecules and produce useful representations of molecular structure, 
which could be used to generate novel and yet drug-like molecules for drug design applications such as virtual screening.

As a proof of concept, we built a DVAE involving transformer layers~\cite{vaswani2017attention} in the encoder and decoder components along with additional preprocessing layers that allowed our model to operate at the character-level 
(rather than on the word-level) to parse SMILES, the textual representations of the input molecules. 
Using SMILES is not necessarily the best option, since the strings are not $100\%$ valid. 
The only property of SMILES that is essential in our approach is that it is a representation of molecules in terms of character strings and hence we believe that DVAEs can be built to operate with alternative character string representations of molecules, 
such as SELFIES~\cite{krenn2020self}.

We trained a compact DVAE with the RBM consisting of two layers of just $128$ units each on a small subset containing almost $200,000$-random molecules from the ChEMBL database of manually curated and biologically active molecules as the training set. 
On classical hardware, the system could be trained with Gibbs sampling. 
We were able to show that the training converged and used the network to generate molecules with the distribution of the basic properties, such as logP, and QED, closely matching those in the training set. 
Simultaneously, the average size of the molecules appeared to increase as the training of the network was progressing. 
There are relatively more harder-to synthesize compounds among the molecules generated by the network. 

Our generative model outputs drug-like molecules and may be deployed on already existing quantum annealing devices (such as D-Wave Advantage processor). 
Training of the same architecture network on the quantum annealer proceeded slower per epoch than on the classical computer, most probably due to noise. 
Nevertheless, the distributions of the molecular properties of generated molecules were sufficiently close to those in the training set or among the molecules generated by classical counterparts Gibbs-75 and 300. 
While certain discrepancies between distributions were present, those results have been computed only after a limited number of training epochs due to the restrictions on public access to the quantum computer. 

Computational drug design applications depend on but are not limited to the generation of novel and synthetically accessible molecules, which is the focus of this work. 
The authors of the original paper~\cite{gomez2018automatic} have already proposed training additional properties, such as the prediction of the binding constant to a particular target on top of the autoencoder loss. 
Although a direct extension of VAE for these tasks may be challenging and require further refinements~\cite{griffiths2020constrained}, 
in such a form, the network could be used in problems involving actual drug design, i.e., for generating of novel compounds binding specific medically relevant targets. 
We did not attempt to demonstrate such a capability. 
However, we have no doubts that DVAE and eventually its hybrid implementations such as QVAE can be appropriately refitted by adding the extra loss.

The RBM could be turned into Quantum Boltzmann Machine (QBM) so that the whole system might be transformed into a Quantum VAE (QVAE,~\cite{khoshaman2018quantum}) and sample from potentially richer non-classical distributions.
Using genuine QBMs should speed up the training of the system ($\mathcal O(\log N)$ vs. $\mathcal O(\sqrt{N})$ with $N$ being the size of the network~\cite{Biamonte2017}). 
There was a demonstration in Ref.~\cite{khoshaman2018quantum}, where ``quantum'' samplers with the non-vanishing transverse fields outperformed DVAE if assessed by metrics achieved at the same number of training cycles (epochs). 
Construction of QVAE with the controllable non-zero transverse field can, in principle, be performed on the existing generation of D-Wave chips. 
However, it would require additional hardware tuning and applying a combination of extra tricks such as reverse anneal schedule, pause-and-quench, etc~\cite{dwavemanual}. 

We demonstrated that a useful VAE can be built and trained to generate drug-like molecules while keeping the size of latent representation small and hence practically attainable on already existing quantum annealing devices. 
We expect that with further developments in the engineering of quantum computing devices, hybrid architectures similar to QVAE would surpass their classical counterparts. 
More specifically, the network architecture proposed in this work may provide the baseline for further refinements required for running genuinely quantum generative models. 
The benefit may be especially large in problems potentially involving rules of quantum chemistry, such as learning efficient representations of molecular structures for applications related to generative chemistry and drug design.

\section{Acknowledgements}
We thank S. Usmanov for help in performing the experiment.
P.O.F and A. K. would like to thank Dr. A. Tarkhov from Gero and Daria Orhunova for fruitful discussion and help with the data analysis. 
We acknowledge online cloud access to the quantum annealing device produced by D-Wave Systems. A.K.F. thanks Russian Science Foundation grant (19-71-10092). P.O.F. and K.~A. are supported by Gero LLC PTE (Singapore).

\section*{Author contributions}
All the authors jointly developed the problem statement and analyzed existing state-of-the-art techniques. 
A.I.G. and A.S.B. implemented the considered methods and performed the simulation of molecules.
All the authors contributed to discussions of the results and writing the manuscript.
P.O.F. and A.K.F. supervised the project. 

\section*{Competing interests}
Owing to the employments and consulting activities of A.I.G., A.S.B., and A.K.F., they have financial interests in the commercial applications of quantum computing. 
P.O.F. and K.A. are employees of (P.O.F. is a stake holder in) Gero LLC PTE and hence may have financial interests in commercial applications of quantum or hybrid AI/ML systems for drug discovery.

\section*{Code availability}
The code that is deemed central to the conclusions is available from the corresponding author upon reasonable request. 
The data that supports the findings of this study (generated molecules) are available at https://doi.org/ 10.5281/zenodo.7827952

\section{Methods}
We proposed and characterized classical and quantum annealer models, which are a combination of Discrete Variational Autoencoder (DVAE) with Restricted Boltzmann Machine (RBM) 
in the latent space~\cite{khoshaman2018quantum, rolfe2016discrete} and the Transformer model~\cite{vaswani2017attention}. 
Original Transformer model was proposed for word-level natural language processing tasks and has encoder-decoder architecture. 
We used original Transformer layers and developed additional preprocessing layers that allowed to process character-level SMILES descriptions of molecules.
We trained the proposed models on a subset of the ChEMBL dataset by optimizing evidence lower bound (ELBO) for DVAE log-likelihood~\cite{khoshaman2018quantum}, 
modified with additional coefficient $\beta$ that multiplies KL divergence term~\cite{yan2020rebalancing}, see Eq.~(\ref{eq:elbo}). 
The sketch of the architecture of our models is illustrated in Fig.~\ref{fig:net-arch}.

Below we in details describe the dataset, the network architecture, the training parameters, and the training schedule of the classical and quantum annealer models.
Also, we describe simpler classical model with continuous latent variables, which we used in the experiment shown in Fig.~\ref{fig:loss}. 

\subsection{Dataset}
We used a subset of molecules from the ChEMBL (release 26) database~\cite{Gaulton2011, Gaulton2017}. 
Our dataset consisted of the $192,000$ structures encoded by SMILES strings of the maximum length of 200 symbols and containing the atoms from the organic subset only (\texttt{B, C, N, O, P, S, F, Cl, Br, I}). 
To focus on the relevant biologically active compounds, we removed salt residuals. Finally, we converted all SMILES into the canonical format with the help of RDKit~\cite{rdkitlib}. 

The processed molecules were randomly assigned into train and validation sets each containing $80\%$ and $20\%$ of all samples ($153600$ and $38400$ molecules), respectively.

\subsection{Training DVAE using Gibbs-sampling}

Molecular SMILES strings are tokenized with the regular expression from Ref.~\cite{schwaller2018found}, which produced $42$ unique tokens.
Standard trainable embedding layer and positional encoding from Ref.~\cite{vaswani2017attention} are used. Our implementation utilized a combination of embedding and positional encoding, in which positional encoding is multiplied by additional correction factor:
\begin{equation}
	\tilde{\mathbf{x}}_{emb} = \sqrt{d_{emb}} \mathbf{x}_{emb}+ \frac{1}{\sqrt{d_{emb}}} \mathbf{pe}
\end{equation}
where $\mathbf{x}_{emb}$ is embedding tensor, $\mathbf{pe}$ is positional encoding tensor and $d_{emb}$ is the dimensionality of the embedding. 
This factor is required to make the proportion between embedding tensor and positional encoding closer to that in the original model~\cite{vaswani2017attention}. 
The dimension of embeddings is a model hyperparameter which was set to $32$. 

We employed a layer of one-dimensional convolutions and a highway layer~\cite{Srivastava2015} as additional preprocessing layers between the embedding layer and the encoder component. 
The convolution layer with $160$ filters and the kernel size equal to $5$ was developed based on Ref.~\cite{banar2020Character}.
We used highway layer, since such layers have been shown to improve the quality of character-level models~\cite{kim2016character, banar2020Character}.

The preprocessed 160-dimensional tensor is passed from the highway layer to the encoder, consisting of the stack of $5$ Transformer encoder layers. 
The width of the feed-forward part of the layers is equal to $320$. Number of heads in Multi-Head attention is $10$. We used GeLU activation~\cite{Hendrycks2016} functions and Dropout with the rate of $0.1$.

Original Transformer encoder layers produce output tensor of variable length. The length of the tensor is equal to the size of the input string. In order to further reduce the dimensionality of the latent space layer in the model, we construct fixed-length tensor from the Transformer encoder output tensor $\mathbf{u}$ by calculating fixed number of vectors from $\mathbf{u}$, which we then concatenate in one tensor.
The first two of these vectors are the vector with index $0$ from Transformer layers output $\mathbf{u}$ and vector equal to the arithmetic mean of all vectors along the length of the tensor $\mathbf{u}$. 
Next, we consider the subsets $S^{m}_{n}$, each consisting of vectors with indices that have the same remainder after division by $n$ for $n = 2, 3, 4, 5$:
$$
	S^{m}_{n} =\{\mathbf{u}_i : i \equiv m \Mod{n} \}, m=0,...,n-1.
$$
For each $S^{m}_{n}$, we compute the arithmetic mean and concatenate all calculated vectors into the fixed-length output tensor.
	
Restricted Boltzmann machine (RBM) is implemented in the latent space as presented in papers~\cite{khoshaman2018quantum, rolfe2016discrete}.
The probability distribution of RBM is
$$
	p_{\boldsymbol{\theta}}(\mathbf{z}) \equiv e^{-E_{\boldsymbol{\theta}}(\mathbf{z})} / Z_{\boldsymbol{\theta}}, \ Z_{\boldsymbol{\theta}} \equiv \sum_{\mathbf{z}} e^{-E_{\boldsymbol{\theta}}(\mathbf{z})},
$$
$$
	E_{\boldsymbol{\theta}}(\mathbf{z}) = \sum_{l} z_l h_l + \sum_{l<m}W_{lm}z_l z_m, \ \mathbf{h}, \mathbf{W} \in \{\boldsymbol{\theta}\},
$$
where $h_l$ are bias weights for units $z_l$ and each $W_{lm}$ is the weight associated with the connection between units $z_l$ and $z_m$. The effective temperature is supposed to be equal to $1.0$ and is not presented in the formulas.
RBM in the proposed model consists of two layers of $128$ units each. 
RBM of this size can be sampled using existing quantum annealing devices. 
It is worth noting that all the units of RBM in DVAE are latent variables and connected to the rest of the model. Hence, there is no distinction between "hidden" and "visible" units as for standalone RBM~\cite{khoshaman2018quantum, rolfe2016discrete}. 

Informal description of the internal working of the model in the latent space is as follows. 
The output of the encoder is the vector of probabilities of the discrete latent variables $z_i$ being equal to $1$, which are conditioned on the input $\mathbf{x}$ of the model. 
These probabilities are sampled to obtain latent binary vector $\mathbf{z}$. Continuous variables $\boldsymbol{\zeta}$ 
are sampled using spike-and-exponential smoothing probability distribution $r(\boldsymbol{\zeta} | \mathbf{z})$ \cite{khoshaman2018quantum}. 
Vector $\boldsymbol{\zeta}$ is used as an input to the decoder module. 
During training, the parameters of the RBM are adjusted in order to memorize the statistics of the binary vectors $\mathbf{z}$ that appear in the latent space. 
The calculation of the gradient of the parameters of the RBM consists of two parts: the so-called "positive" and "negative" phases. 
The "positive" phase is calculated using backpropagation algorithm after application of the reparameterization trick, which is used to avoid calculating of derivatives over random variables. 
The "negative" phase of the gradient is estimated using sampling from the RBM distribution. 

For molecule reconstruction or generation of similar molecules to a given molecule, the preprocessed SMILES description of the given molecule is passed to the input of the encoder and the whole model is executed.
An example of molecule reconstruction and generation of similar molecules is depicted in Fig..~\ref{fig:net-arch}.
For generation of an entirely new molecule the encoder is not used, the trained RBM is sampled to obtain latent binary vector $\mathbf{z}$. 
This vector is then used to calculate the latent vector of continuous variables $\boldsymbol{\zeta}$, which is given as an input to the decoder. 
Table~\ref{table:metrics} and Fig.~\ref{fig:comparison} show results for newly generated molecules.

RBM is sampled by performing $30$ steps of Gibbs updates using persistent contrastive divergence (PCD)~\cite{tieleman2008training}. 

The decoder works in two modes: training and inference (generation). In the inference mode decoder uses preprocessing layers. 
The main part of data processing in both training and inference modes of decoder consists of Transformer decoder layers. 
Altogether, we used $5$ Transformer decoder layers of the size $d_{model}=160$ (GeLU activation, dropout = $0.1$). 
The width of the feed-forward part of the layers was equal to $320$, the number of heads in Multi-Head attention was $10$. 

To train the model, we used the rebalanced objective function, in which the KL divergence term is multiplied by the additional coefficient $\beta = 0.1$~\cite{yan2020rebalancing} to avoid posterior collapse problem, and employed the Adam optimizer.

In contrast to the original Transformer model, we used a different learning rate schedule: we trained the model for $300$ epochs using MultiStep learning rate schedule with the initial learning rate equal to $6\times10^{-5}$. The learning rate was subsequently reduced by the factor of $0.5$ at points corresponding to $50\%$, $75\%$, and $95\%$ of the length of the training process. 

For estimation of the logarithm of the partition function of Boltzmann distribution, we used annealed importance sampling (AIS) algorithm~\cite{Neal2001AnnealedIS} 
during the evaluation of the model at the end of each epoch using $10$ intermediate distributions and $500$ samples. 

Due to resources constraints, we did not have a chance to optimize the hyperparameters or too many architectural variants of the model. 
The presented variant of the network just worked and can be considered the first step towards a real and effective solution.

\subsection{Training DVAE on a quantum annealer}
We used exactly the same network architecture on the quantum annealer with the only difference from the classical case being that the RBM in the latent space was sampled using quantum annealer (D-Wave Advantage processor). Also, the quantum model was trained during $75$ epochs with constant learning rate equal to $6\times10^{-5}$.

For estimating the logarithm of the partition function of the Boltzmann distribution during the evaluation of the model, 
we used a different version of annealed importance sampling (AIS; see Ref.~\cite{Ulanov2019QuantuminspiredAA}) with the same parameters as in the classical case.

\subsection{Training model with continuous latent variables}
The model with continuous variables in the latent space has similar architecture to the discrete one but is smaller in size. 
The latent space contains $32+32$ normally distributed continuous random variables.

The preprocessing convolution layer consists of $100$ filters with kernel size equal to $5$. 
The encoder/decoder consists of $2$ Transformer encoder/decoder layers with the width of feed-forward part equal to $200$.

The fixed length tensor is calculated in the similar way as in the discrete model. 
The model is trained using the same initial learning rate and learning rate schedule as in the discrete case. 

\subsection{Calculation of molecular similarity with fingerprint}

Fingerprints for each molecule are generated using a default function \texttt{RDKFingerprint} in RDKit~\cite{rdkitlib}. 
This algorithm produces topological fingerprint represented by a bit vector with the size of 2048 bits. 
The Tanimoto similarity is known as a reasonable metric for matching molecules sharing similar fragments~\cite{Bajusz2015} and is defined for two fingerprints $a,b$ as:
\begin{equation}
  T(a,b)=\frac{C}{A+B-C}
\end{equation}
where $C$ is a number of common non-zero bits in $a$ and $b$; $A$ and $B$ are numbers of non-zero bits that are present in $a$ and $b$, respectively. 
The Tanimoto distance could be defined as $D(a,b)=1-T(a,b)$. 
From the definition, it follows that completely similar molecules (shared identical set of fragments) have Tanimoto similarity equal to 1, while dissimilar molecules (no common fragments) have $T=0$.

\bibliographystyle{naturemag}
\bibliography{QVA.bib}

\end{document}